\documentclass[12pt]{article}
\usepackage{amsfonts}
\newsavebox{\uuunit}
\sbox{\uuunit}
    {\setlength{\unitlength}{0.825em}
     \begin{picture}(0.6,0.7)
        \thinlines
        \put(0,0){\line(1,0){0.5}}
        \put(0.15,0){\line(0,1){0.7}}
        \put(0.35,0){\line(0,1){0.8}}
       \multiput(0.3,0.8)(-0.04,-0.02){12}{\rule{0.5pt}{0.5pt}}
     \end {picture}}

\begin{document}
\begin{flushright}
FTUV-06-0710 \hskip .4cm IFIC/06-27\\
July 10, 2006\\
JHEP 09 (2006) 009
\end{flushright}
\vspace{.5cm}

\begin{center}
 {\bf \large On the absence of BPS preonic solutions in IIA and
IIB supergravities}

\bigskip

{ Igor A. Bandos$^{1,2}$, Jos\'e A. de Azc\'arraga $^1$ and Oscar
Varela$^1$}

\bigskip

$^1$ {\it Department of Theoretical Physics, Valencia University,
and IFIC (CSIC-UVEG), 46100-Burjassot (Valencia), Spain}

$^2${\small\it Institute for Theoretical Physics, NSC KIPT,
 61108 Kharkov, Ukraine}


\end{center}

\begin{center}
{\bf Abstract}
\end{center}
{\small We consider the present absence of $\nu=31/32$
supersymmetric solutions in supergravity {\it i.e.}, of solutions
describing BPS preons. A recent result indicates that (bosonic)
BPS preonic solutions do not exist in type IIB supergravity. We
reconsider this analysis by using the $G$-frame method, extend it
to the IIA supergravity case, and show that there are no (bosonic)
preonic solutions for type IIA either. For the classical $D=11$
supergravity no conclusion can be drawn yet, although the negative
IIA results permit establishing the conditions that preonic
solutions would have to satisfy. For supergravities with `stringy'
$(\alpha^\prime)^3$-corrections, the existence of BPS preonic
solutions remains fully open.}


{\small \tableofcontents}

\section{Introduction}

 It has been argued in a very recent paper \cite{GGPR06} that
purely bosonic solutions preserving $31$ out of $32$
supersymmetries, hence describing {\it BPS preon} states
\cite{BPS01}, do not exist for IIB supergravity. Using the moving
$G$-frame method of \cite{BPS03} (Sec. 1.2), we rederive this
result here (Sec. 2). Then, we apply the same technique to the IIA
case and also show that preonic solutions do not exist in type IIA
supergravity (Sec. 3). Nevertheless, the concluded absence of
preonic solutions could be modified if the `stringy'
$(\alpha^\prime)^3$-corrections to the dilatino transformation
rule were made explicit and taken into account (Sec. 5).

For $D=11$ supergravity, the existence of BPS preonic solutions is
not ruled out even at the classical level ({\it i.e.}, ignoring
$(\alpha^\prime)^3$-corrections), although the above negative
results for type IIA supergravity already set strong restrictions
(Sec. 4) to be satisfied by these solutions.

\subsection{Basic notions and notation}

In eleven-dimensional supergravity \cite{CJS78} the only fermionic
field is the gravitino, $\check{\psi}^{\;\check{\alpha}}=
dx^{\check \mu} \check{\psi}_{\check{\mu}}^{\;\check{\alpha}} =
dx^{\mu}\check{\psi}_{{\mu}}^{\;\check{\alpha}} +
dx^{\#}\check{\psi}_{{\#}}^{\;\check{\alpha}} $ (${\check \mu} =
(\mu;\#), \; \mu=0,1,\dots,9$). In contrast, the ten-dimensional
type II supergravities \cite{IIBsugra,IIAsugra} contain, in
addition to two sixteen-component `spin 3/2' gravitini, two `spin
1/2' dilatini fields $\check{\chi}_{\check{\alpha}}$. We use the
czek superscript $\check{\alpha}$ to denote the type II indices of
the $32$-component reducible spinors. In the IIB case
$\check{\alpha}$ is the double index $\check{\alpha}=(\alpha, I)$,
where $I=1,2\,$ labels the two Majorana-Weyl (MW) spinors of the
same chirality and $\alpha=1, \ldots 16$. In the IIA case, where
both chiralities are present, $\check{\alpha}$ denotes the
Majorana spinor index and thus $\check{\alpha}=1,\ldots , 32$.

In particular, for the dilatino of type IIA supergravity we write
\begin{eqnarray}\label{IIADIL}
\hbox{IIA} \; :\qquad \qquad \check{\chi}^{\check{\alpha}}:=
({\chi}^{{\alpha}1}\; , \; {\chi}^2_{{\alpha}})\; , \qquad
\check{\alpha}=1, \ldots ,32\; , \quad \alpha=1, \ldots , 16 \; .
\end{eqnarray}
while in type IIB supergravity the $32$-component dilatino field
decomposes into two MW spinors of the same chirality,
\begin{eqnarray}\label{IIBDIL}
\hbox{IIB} \; : \qquad \check{\chi}_{\check{\alpha}}:=
({\chi}^1_{{\alpha}}\; , \; {\chi}^2_{{\alpha}})\; , \quad
\check{\alpha}=(\alpha\,, I)\; , \quad  I=1,2 \; , \quad \alpha=1,
\ldots , 16 \; . \quad
\end{eqnarray}
Notice that in the IIB case the position of the index
$\check{\alpha}$ cannot be changed since the two MW spinors are of
the same chirality and there is no $16\times16$ charge conjugation
matrix in the MW spinor representation. In contrast, in type IIA a
$32\times 32$ charge conjugation matrix exists; it is
anti-diagonal in the Weyl-like realization used here and exchanges
the 1 and 2 MW components in (\ref{IIADIL}).

In this condensed 32-component notation, the supersymmetry
transformation rules for the gravitini and dilatini fermionic
fields can be written in compact form for both IIA and IIB cases
as
\begin{eqnarray}
\label{susyDIL}
\delta_{susy} \check{\psi}_a^{\check{\alpha}}= {\cal
D}_a\check{\varepsilon}{}^{\,\check{\alpha}}:= D_a
\check{\varepsilon}{}^{\,\check{\alpha}}-
\check{\varepsilon}{}^{\check{\beta}} {\check t}_{a\,
\check{\beta}}{}^{\check{\alpha}}\; , \qquad \delta_{susy}
\check{\chi} = \check{\varepsilon} M\; ,  \qquad
\end{eqnarray}
where $D=d-\omega$ is the Lorentz covariant derivative and ${\cal
D}=D-{\check t}$ is the generalized covariant derivative which
includes, besides the (suitable) spin connection $\omega:= {1\over
4}\omega^{ab}{\check\Gamma}_{ab}$, the additional tensorial IIA or
IIB ${\check t}$ contributions. The transformation rules for the
dilatino are algebraic and are characterized by a $32\times 32$
matrix $M_{\check{\beta}}{}^{\check{\alpha}}$. The form that this
matrix takes will be crucial for the discussion below.

In the IIA case, and ignoring inessential bilinear fermionic
contributions, the terms in $\delta_{susy} \check{\chi}$ (Eq.
(\ref{susyDIL}), see {\it e.g.} \cite{IIAsugra} and \cite{MMS03}
and refs therein) are determined by the matrix
\begin{eqnarray}\label{M=IIA}
\hbox{IIA} \; : \qquad M_{\check{\beta}}{}^{\check{\alpha}}=
\left(\matrix{ {3\over 8} e^{\Phi} R\!\!\!\!/^{(2)}  + {1\over 8}
R\!\!\!\!/^{(4)} & {1\over 2} \partial\!\!\!\!/\Phi- {1\over 4}
H\!\!\!\!/^{(3)}  \cr {1\over 2} \tilde{\partial\!\!\!\!/}\Phi +
{1\over 4} \tilde{H}\!\!\!\!/^{(3)}  & - {3\over 8} e^{\Phi}
\tilde{R}\!\!\!\!/^{(2)}  + {1\over 8} \tilde{R}\!\!\!\!/^{(4)}
}\right) \; .  \qquad
\end{eqnarray}
in terms of all the possible IIA fluxes (on-shell field
strengths), namely\footnote{$\sigma^a= \sigma^a_{\alpha\beta}$,
${\tilde \sigma}^a= {\sigma^a}^{\alpha\beta}$, $a=0,1,\dots,9\,$;
$\;\sigma^a{\tilde\sigma}^b+\sigma^b{\tilde\sigma}^a= 2 \eta^{ab}
={\tilde\sigma}^a\sigma^b+{\tilde\sigma}^b \sigma^a$. The sigma
matrices with one and five (three) vector indices are symmetric
(antisymmetric) with respect to the spinor ones. The transposition
of untilded sigma matrices with four and two vector indices,
respectively, converts them into the corresponding tilded and
minus tilded ones.},
\begin{eqnarray}
\label{fluxesIIA}
 R_2:=dC_1\; , \qquad R_4:=dC_3-C_1\wedge H_3\; , \qquad H_3:=dB_2 \; \qquad and
\qquad d\Phi \; :  \qquad
\end{eqnarray}
\begin{eqnarray}
\label{fl-NSNS}  \cases{ H\!\!\!\!/^{(3)}= {1\over 3!}
H_{abc}\sigma^{abc} \; , \qquad \sigma^{abc}:=
({\sigma}^{[a}\tilde{\sigma}^{b}\sigma^{c]})_{\alpha\beta}\; , \cr
\tilde{H}\!\!\!\!/^{(3)}= {1\over 3!} H_{abc}\tilde{\sigma}^{abc}
\; , \qquad \tilde{\sigma}^{abc}:= (\tilde{\sigma}^{[a}\sigma^
{b}\tilde{\sigma}^{c]})^{\alpha\beta}}
 \; ,  \qquad
\cases{ \partial\!\!\!\!/\Phi := \partial_a\Phi
\sigma^a_{\alpha\beta}\; , \cr \tilde{\partial}\!\!\!\!/\Phi :=
\partial_a\Phi \tilde{\sigma}^{a\, \alpha\beta}\; ,}\qquad
\\ \label{fl-RRIIA} \cases{ R\!\!\!\!/^{(2)}:= {1\over 2!}
R_{ab}(\sigma^{ab})= - \tilde{R}\!\!\!\!/^{(2)}{}^T \; , \qquad
\sigma^{ab}:= (\sigma^{[a}\tilde{\sigma}^{b]}){}_\alpha{}^\beta \;
, \qquad \tilde{\sigma}^{ab}:=
(\tilde{\sigma}^{[a}\sigma^{b]}){}^\beta{}_\alpha \; ,\cr
R\!\!\!\!/^{(4)}= {1\over 4!}
R_{abcd}\sigma^{abcd}{}_\alpha{}^\beta =
(\tilde{R}\!\!\!\!/^{(4)}){}^\beta{}_\alpha\; , \qquad
\sigma^{abcd}:=
(\sigma^{[a}\tilde{\sigma}^{b}{\sigma}^{c}\tilde{\sigma}^{d]}){}_\alpha{}^\beta
\; . \quad }
\end{eqnarray}

The type IIB matrix  $M$, in contrast, is given by (see
\cite{IIBsugra})
\begin{eqnarray}\label{M=IIB}
\hbox{IIB} \; : \qquad M_{\check{\beta}\check{\alpha}}=
\left(\matrix{ {1\over 2} {\partial\!\!\!\!/}\Phi + {1\over 4}
{H}\!\!\!\!/^{(3)} & - {1\over 2} e^{\!^{\Phi}} R\!\!\!\!/^{(1)} +
{1\over 4} e^{^{{1\over 2}\Phi}} R\!\!\!\!/^{(3)} \cr
 {1\over 2}
e^{\!^{\Phi}} R\!\!\!\!/^{(1)}  + {1\over 4} e^{^{{1\over 2}\Phi}}
R\!\!\!\!/^{(3)} & {1\over 2} {\partial\!\!\!\!/}\Phi - {1\over 4}
{H}\!\!\!\!/^{(3)} }\right) \; , \qquad
\end{eqnarray}
and involves the one-form and the three-form fluxes of type IIB
supergravity,
\begin{eqnarray}\label{fl-1,3IIB}
  \;\; R_1:=dC_0\; , \quad R_3:=dC_2-C_0H_3\; ,  \qquad
H_3:=dB_2 \; \qquad and \qquad d\Phi \; ,   \qquad
\end{eqnarray}
but not the self dual five-form flux $R_5$,
\begin{eqnarray}\label{fl-5IIB}
R_5:=dC_4-C_2\wedge H_3\; , \qquad  R_5=*R_5 \; \Leftrightarrow
\cases{ R\!\!\!\!/^{(5)} =0 \; , \cr
\tilde{R}\!\!\!\!/^{(5)}\not=0 \; . }
\end{eqnarray}

When only purely bosonic solutions are considered,
$\check{\psi}=0$, $\check{\chi}=0$, the parameter associated with
the  preserved supersymmetry obeys a differential equation and an
algebraic one, namely ${\cal D}\check{\varepsilon}=0$ and
$\check{\varepsilon}M=0$. Usually, to describe a solution
preserving $k$ supersymmetries (a $\nu=k/32$ state), one uses $k$
independent bosonic Killing spinors $\epsilon_I^{\check\alpha}$
($I=1,\ldots , k$,
$\check{\varepsilon}=\kappa^I\epsilon_I^{\check\alpha}$ with
arbitrary constant fermionic $\kappa^I$) that satisfy the
following differential (from $\delta_{susy}
\check{\psi}_a^{\check{\alpha}}=0$) and algebraic (from
$\delta_{susy} \check{\chi}=0$ ) Killing equations
\begin{eqnarray}\label{KEqGR}
 && {\cal D}\check{\epsilon}_I := D
\check{\epsilon}_I - \check{\epsilon}_I {\check t}\, = 0\; ,
\qquad
\\ \label{KEqDIL} && \check{\epsilon}_I \, M\;
=0 \qquad (I=1, \ldots , k)\; , \qquad
\end{eqnarray}
which guarantee that the solution remains bosonic and hence
invariant after a gravitino and dilatino supersymmetry
transformation.

 The conclusion of \cite{GGPR06} on the absence of a preonic
solution of type IIB supergravity is based on the algebraic
equation (\ref{KEqDIL}) and uses (\ref{KEqGR}) to close the
argument. We now recover this result below by using the moving
$G$-frame method of \cite{BPS03}.

\subsection{The moving $G$-frame method and preonic spinors}

A preonic state \cite{BPS01} preserves all supersymmetries but
one; it is a $\nu=31/32$ supersymmetric BPS state. As a result,
 it can be characterized by one bosonic spinor
$\check{\lambda}_{\check{\alpha}}$ orthogonal to all the $31$
bosonic Killing spinors $\check{\epsilon}_I^{\;\check{\alpha}}$ in
(\ref{KEqGR}),
\begin{eqnarray}
\label{PreonL} \check{\epsilon}_I \, \check{\lambda} = \;
\check{\epsilon}_I^{\; \check{\alpha}}
\check{\lambda}_{\check{\alpha}} =0 \; , \qquad I=1, \ldots , 31\;
. \qquad
\end{eqnarray}
As it was noticed in  \cite{BPS03}, when the  generalized holonomy
group of supergravity  \cite{Duff+Stelle91,Duff03} is a subgroup
of $SL(32,\mathbb{R})$ (which is the case for both $D=11$
\cite{Hull03} and type II $D=10$ supergravities \cite{P+T03}), the
spinor characterizing a BPS preonic state obeys the differential
equation
\begin{eqnarray}
\label{Dprl} {\cal D} \check{\lambda} := D \check{\lambda} +
\check{t} \check{\lambda} =0 \;  , \qquad
\end{eqnarray}
where ${\check t}$ is the same tensorial part of the generalized
connection in Eqs. (\ref{KEqGR}) and (\ref{susyDIL}). Notice that
if ${\check t}\not= 0$ (the case of non-vanishing fluxes), Eq.
(\ref{Dprl}) is not equivalent to the Killing equation
(\ref{KEqGR}) even for the type IIA case where the $32\times 32$
charge conjugation matrix does exist.

Applied to the present problem, the moving $G$-frame method
\cite{BPS03} implies that  Eq. (\ref{KEqDIL}), looked at as an
equation for the matrix $M$, is solved when $k=31$ by
\begin{eqnarray}
\label{KEqDIL31} M = \check{\lambda} \otimes \check{s} \qquad i.e.
\qquad \cases{IIA\; : \; M_{\check{\beta}}{}^{\check{\alpha}}=
\check{\lambda}_{\check{\beta}} \; \check{s}^{\check{\alpha}} \; ,
\cr IIB\; : \; M_{\check{\beta}\check{\alpha}}=
\check{\lambda}_{\check{\beta}} \; \check{s}_{\check{\alpha}} \;
,} \qquad
\end{eqnarray}
where  $\check{s}_{\check{\alpha}}$ is a certain spinor. The
algebraic structure of the matrix $M$ implies a series of
restrictions on the preonic spinor
$\check{\lambda}_{\check{\beta}}$. At the same time, Eq.
(\ref{KEqDIL31}) imposes a series of restrictions on the fluxes
involved in the matrix $M$.

Eq. (\ref{KEqDIL31}) will be the basic equation in our analysis of
the absence of preons among the bosonic solutions of type II
supergravities.

\section{Absence of preons in type IIB supergravity}

In the  type IIB case the matrix $M$ has the form of Eq.
(\ref{M=IIB}), and  Eq. (\ref{KEqDIL31})  implies the following
relations for the one- and three-form fluxes
\begin{eqnarray}
\label{Eq-IIB-pr1}
\begin{matrix}
{ & {1\over 2} \partial\!\!\!\! / \Phi  + {1\over
4}H\!\!\!\!/^{(3)} = \; \lambda^1_\alpha s_\beta^1  \; , &
\hbox{(\ref{Eq-IIB-pr1}a)} & \cr & - {1\over 2} e^{\!^{\Phi}}
R\!\!\!\!/^{(1)}  + {1\over 4} e^{^{{1\over 2}\Phi}}
R\!\!\!\!/^{(3)}= \lambda^1_\alpha s_\beta^2 \; , &
\hbox{(\ref{Eq-IIB-pr1}b)} & \cr
 & + {1\over 2}
e^{\!^{\Phi}} R\!\!\!\!/^{(1)}  + {1\over 4} e^{^{{1\over 2}\Phi}}
R\!\!\!\!/^{(3)} = \; \lambda^2_\alpha s_\beta^1  \; , &
\hbox{(\ref{Eq-IIB-pr1}c)} & \cr \qquad  & {1\over 2}
{\partial\!\!\!\!/}\Phi - {1\over 4} {H}\!\!\!\!/^{(3)} = \;
\lambda^2_\alpha s_\beta^2 \; . & \hbox{(\ref{Eq-IIB-pr1}d)} &
\qquad }  \end{matrix}
\end{eqnarray}
These fluxes then can be expressed through the IIB preonic spinor
$\check{\lambda}_{\check{\alpha}}:= (\lambda^1_{{\alpha}}\; , \;
{\lambda}^2_{{\alpha}})$ and an arbitrary spinor
$\check{s}_{\check{\beta}}:= ({s}^1_{{\beta}}\; , \;
{s}^2_{{\beta}})$. Furthermore, the consistency of Eqs.
(\ref{Eq-IIB-pr1}) imposes a set of algebraic equations on these
two spinors. They follow from the fact that the fluxes enter into
Eqs. (\ref{Eq-IIB-pr1}) through matrices which possess definite
symmetry properties,
\begin{eqnarray}\label{IIBsym-asym}
(\partial\!\!\!\! / \Phi)^T = +\partial\!\!\!\! / \Phi \; , \qquad
(H\!\!\!\!/^{(3)})^T=- H\!\!\!\!/^{(3)}  ,   \qquad
(R\!\!\!\!/^{(3)})^T=- R\!\!\!\!/^{(3)}  ,   \qquad
(R\!\!\!\!/^{(1)})^T=+R\!\!\!\!/^{(1)}  .   \qquad
\end{eqnarray}
These lead to the algebraic constraints
\begin{eqnarray}
\label{Eq-IIBls}
\begin{matrix}
{ \hbox{(a)} \qquad & \lambda^1_{[\alpha} s_{\beta]}^1 +
\lambda^2_{[\alpha} s_{\beta]}^2 =0  \; , \qquad  & \qquad
\hbox{(b)} \qquad  &  - \lambda^1_{[\alpha} s_{\beta]}^2 +
\lambda^2_{[\alpha} s_{\beta]}^1 =0  \; ,   & \cr
 \hbox{(c)} \qquad & \lambda^1_{(\alpha} s_{\beta )}^1 -  \lambda^2_{(\alpha} s_{\beta)}^2 =0  \; ,
 \qquad  & \qquad
\hbox{(d)} \qquad &  \lambda^1_{(\alpha} s_{\beta )}^2 +
\lambda^2_{(\alpha} s_{\beta )}^1 =0  \; .  & }
\end{matrix}
\end{eqnarray}

A straightforward algebra shows that Eqs. (\ref{Eq-IIBls}) have
only trivial solutions. This means that either the preonic or the
auxiliary spinor is zero,
\begin{eqnarray} \label{lorS=0IIB}
\hbox{IIB} \; :  \qquad \lambda^1_{\alpha}=\lambda^2_{\alpha}=0
\qquad or \qquad
 s_{\beta}^1 = s_{\beta}^2=0 \; . \qquad
\end{eqnarray}
In both cases the matrix  $M=0$ and, hence, all the fluxes except
the five-form flux (Eq.(\ref{fl-5IIB})) are equal to zero,
$R_1=d\Phi=R_1=R_3=0$. Nevertheless, the fact that the solution
$s_{\beta}^1=s_{\beta}^2=0$ of (\ref{Eq-IIBls}) allows for a
non-vanishing preonic spinor
$(\lambda^1_{\alpha},\lambda^2_{\alpha})$ might give hope, at this
stage, of finding a  nontrivial and unique solution ${\check
\lambda}$ to Eq. (\ref{Dprl}) and $k=31$ solutions
$\check{\epsilon}_I$ for Eq. (\ref{KEqGR}). This possibility is
ruled out by looking at Eq. (\ref{KEqGR}). For simplicity let us
begin by discussing Eq. (\ref{Dprl}). When only the five-form flux
is non-vanishing, Eq. (\ref{Dprl}) would acquire the relatively
simple form of
\begin{eqnarray} \label{KEq-l12-IIB}
 \qquad  R_1=R_3=H_3=d\Phi=0 \;: \quad \qquad \left\{
\begin{matrix} {D_b\lambda^1_{\alpha}= - {1\over 16} (\sigma_b
R\!\!\!\!/^{(5)})_{\alpha}{}^{\beta}\lambda^2_{\beta} \; , \cr
D_b\lambda^2_{\alpha}= \; \; {1\over 16} (\sigma_b
R\!\!\!\!/^{(5)})_{\alpha}{}^{\beta}\lambda^1_{\beta}\; . }
\end{matrix}\right.
\end{eqnarray}
Now one observes that, if $(\lambda^1_{\alpha},
\lambda^2_{\alpha})$ is a solution of Eq.(\ref{KEq-l12-IIB}),
$(-\lambda^2_{\alpha}, \lambda^1_{\alpha})$ provides another one.
As a result, the number of solutions of Eqs. (\ref{KEq-l12-IIB})
is always even. The same is true of the Killing equation
(\ref{KEqGR}) since it has the same structure. Hence with
vanishing one- and three-form fluxes one can only have an even
number of preserved supersymmetries. These might include
two-preonic solutions (preserving $30$ supersymmetries) besides
those preserving all $32$ supersymmetries, but not a preonic
solution. The authors of \cite{GGPR06} then concluded that preonic
solutions do not exist for type IIB supergravity.

We now apply our $G$-frame approach, used above to rederive the
IIB result of \cite{GGPR06}, to show that preonic solutions are
also absent in type IIA supergravity.

\section{Absence of preons in type IIA supergravity}

 The crucial point is that in the  IIA case the matrix $M$, Eq.
(\ref{M=IIA}), receives contributions from all IIA fluxes, Eq.
(\ref{fluxesIIA}). Hence if $M$ is zero, all IIA fluxes are zero,
the generalized covariant derivative ${\cal D}$ becomes the
Lorentz covariant derivative $D$ and the generalized holonomy
group reduces to $SO(1,9)$, for which the number of possible
preserved supersymmetries is known (see \cite{Duff03, Bry00}).

As we shall see presently, $M$ is indeed zero if we assume the
existence of $31$ Killing spinors. In type IIA supergravity the
preonic $\check{\lambda}_{\check{\alpha}}$ and auxiliary
$\check{s}^{\check{\alpha}}$ spinors are $32$-component $D=10$
Majorana spinors,
\begin{eqnarray}\label{IIAprl}
\hbox{IIA} \; : \qquad \check{\lambda}_{\check{\alpha}}:=
({\lambda}^1_{{\alpha}}\; , \; {\lambda}^{{\alpha}2})\; , \qquad
\check{s}^{\check{\alpha}}:= ({s}^{{\alpha}1}\; , \;
{s}^2_{{\alpha}})\; ,  \qquad \alpha=1,\ldots, 16\; .  \qquad
\end{eqnarray}
Eq.(\ref{KEqDIL31}) can be split into four equations for the
($16\times16$)-component blocks
\begin{eqnarray} \label{M=lsIIA}
\begin{matrix} { & {3\over
8} e^{\!^{\Phi}} R\!\!\!\!/^{(2)} + {1\over 8} R\!\!\!\! / ^{(4)}=
{\lambda}^1_{{\alpha}} {s}^{{\beta}1}\; ,  & \quad
\hbox{(\ref{M=lsIIA}a)}  \quad  &
 {1\over 2} \partial\!\!\!\! / \Phi - {1\over 4} H\!\!\!\!/^{(3)} =
 {\lambda}^1_{{\alpha}} {s}^2_{\beta}\; , & \quad
\hbox{(\ref{M=lsIIA}b)}   & \cr & {1\over 2} \tilde{
\partial\!\!\!\! / } \Phi + {1\over 4}\tilde{ H\!\!\!\!/^{(3)} }=
{\lambda}^{{\alpha}2} {s}^{{\beta}1}\; ,  & \quad
\hbox{(\ref{M=lsIIA}c)}  \quad  & -{3\over 8} e^{\!^{\Phi}}
\tilde{ R\!\!\!\!/ }^{(2)} + {1\over 8}
 \tilde{R\!\!\!\! / }^{(4)}=
{\lambda}^{{\alpha}2} {s}^2_{{\beta}}\; .  & \quad
\hbox{(\ref{M=lsIIA}d)}   & }
\end{matrix}  \;
\end{eqnarray}
We now notice that $\tilde{ R\!\!\!\!/ }^{(2)}= - ({R\!\!\!\!/
}^{(2)})^T$, $\tilde{ R\!\!\!\!/ }^{(4)}= +( {R\!\!\!\!/
}^{(4)})^T$ and that, accordingly, the {\it l.h.s.}'s of Eqs.
(\ref{M=lsIIA}a) and (\ref{M=lsIIA}d) are equal among themselves.
Hence, the {\it r.h.s.}'s of these equations are also equal,
${\lambda}^1_{{\alpha}} {s}^{{\beta}1}= {\lambda}^{{\beta}2}
{s}^2_{{\alpha}}$. This equation identifies the components of
$\check{\lambda} $ and $\check{s}$ up to a factor $a$,
\begin{eqnarray}\label{lbd=as}
 {s}^{{\alpha}1} = a {\lambda}^{{\alpha}2}\; , \qquad
{s}^2_{{\alpha}}= a {\lambda}^1_{{\alpha}}\;  . \qquad
\end{eqnarray}
Then, decomposing Eq. (\ref{M=lsIIA}a) or (\ref{M=lsIIA}d) into
their irreducible parts ({\it i.e.}, identifying the coefficients
of the matrices $\sigma^{ab}{}_{\alpha}{}^{\beta}$,
$\sigma^{abcd}{}_{\alpha}{}^{\beta}$ {\it and}
$\delta_{\alpha}{}^{\beta}$, one finds the expressions for the RR
fluxes in terms of preonic spinors as well as an orthogonality
condition between $\lambda^1$ and $\lambda^2$,
\begin{eqnarray}\label{R2=...}
e^\Phi R_{ab}= - {a\over 6} \lambda^2 \sigma_{ab}\lambda^1 \; ,
\qquad
 R_{abcd}= {a\over 2} \lambda^2 \sigma_{abcd}\lambda^1 \; , \qquad
 {\lambda}^{{\alpha}2}
{\lambda}^1_{{\alpha}}\; =0   . \qquad
\end{eqnarray}

Substituting (\ref{lbd=as}) for the $s$ spinors in
(\ref{M=lsIIA}b) and (\ref{M=lsIIA}c), these equations can be
rewritten in the form
\begin{eqnarray} \label{dP=all}
 {1\over 2} \partial\!\!\!\! / \Phi - {1\over 4} H\!\!\!\!/^{(3)}
 = a{\lambda}^1_{{\alpha}} {\lambda}^1_{\beta}\; ,
 \qquad
\hbox{(\ref{dP=all}a)}   \qquad  {1\over 2} \tilde{
\partial\!\!\!\! / } \Phi + {1\over 4}\tilde{ H\!\!\!\!/^{(3)} }=
a{\lambda}^{{\alpha}2} {\lambda}^{{\beta}2}\; . \qquad
\hbox{(\ref{dP=all}b)}   \qquad
\end{eqnarray}
The {\it r.h.s.}'s of Eqs. (\ref{dP=all}) are symmetric, while the
{\it l.h.s.}'s contain the antisymmetric matrices $ H\!\!\!\!/ =-
(H\!\!\!\!/ )^T$ and $\tilde{ H\!\!\!\!/ }=-(\tilde{ H\!\!\!\!/
})^T$ which, hence, should be equal to zero. This implies the
vanishing of  the NS-NS flux $H_3$ for a hypothetical preonic
solution of type IIA supergravity, $ H_{abc}=0$. Then one arrives
at
\begin{eqnarray} \label{dP=allV}
 {1\over 2} \sigma^a_{{\alpha}\beta}D_a \Phi =
a{\lambda}^1_{{\alpha}} {\lambda}^1_{\beta}\; ,
 \qquad
 {1\over 2} \tilde{\sigma
}{}^{a{\alpha}\beta} D_a  \Phi = a{\lambda}^{{\alpha}2}
{\lambda}^{{\beta}2}\; . \qquad
\end{eqnarray}
Since we are in ten dimensions these equations imply, besides
$D_a\Phi \sim \lambda^1 {\tilde \sigma}_a\lambda^1$,
\begin{eqnarray} \label{ls5l=0}
 a {\lambda}^1 \tilde{\sigma}^{a_1\ldots a_5} {\lambda}^1=0 \; ,
 \qquad
a{\lambda}^{2}{\sigma}^{a_1\ldots a_5} {\lambda}^{2}=0 \; . \qquad
\end{eqnarray}

Eqs. (\ref{dP=allV}) or  (\ref{ls5l=0}) imply the {\it absence of
BPS preons among the bosonic solutions of type IIA supergravity}.
Indeed, for non-vanishing $a$ ($a\not=0$) Eqs.(\ref{ls5l=0}) have
only trivial\footnote{A simple way to prove it from Eq.
(\ref{dP=allV}) is to notice that this equation implies $D_a\Phi
\propto \lambda^1\tilde{\sigma}_a\lambda^1$ and that, hence,
$D_a\Phi$ is a light-like ten-vector, $D_a\Phi D^a\Phi=0$. Then
one may choose the Lorentz frame where $D_a\Phi \propto
(1,0,\ldots , 0, \pm 1)$; in it, $D_a\Phi \sigma^a_{\alpha\beta}
\propto (\sigma^0_{\alpha\beta} \pm \sigma^{9}_{\alpha\beta}) =
2\sum_{p}\delta^p_{\alpha}\delta^p_{\beta}$, where $p=1, \ldots ,
8$. In this frame, the first equation in (\ref{dP=allV}) reads
$D_0 \Phi\sum_{p} \delta^p_{\alpha}\delta^p_{\beta} =
a{\lambda}^1_{{\alpha}} {\lambda}^1_{\beta}\,$, which immediately
implies that $a\not=0$ is only possible if half of the sixteen
components of ${\lambda}^1_{\beta}$ are zero,
${\lambda}^1_{\beta}={\lambda}_q \delta^q_{\beta}$. Taking this in
account, the above equation reduces to $D_0 \Phi \delta_{qp}=
a\lambda_q \, \lambda_p $ with $p,q=1, \ldots ,8 \,$, which for
$a\not=0$ only admits the trivial solution
$\lambda^1=0=\lambda^2$. } solutions, $\lambda^1=0=\lambda^2$.
This may correspond to the case of a fully supersymmetric solution
of supergravity (preserving the $32$ supersymmetries), but not to
a preonic one. The other possibility, $a=0$, also implies the
vanishing of the $M$ matrix (\ref{M=IIA}) and hence of all type
IIA supergravity fluxes, $R_2=0=R_4$, $H_3=0=d\Phi$, and thus the
generalized connection in the Killing equation (\ref{KEqGR})
reduces to the spin-connection, ${\cal D}=D$. In such a case it is
known (see \cite{Duff03, Bry00}) that the Killing spinor equation
$D\check{\epsilon}=0$ may have either $32$ or up to $16$
solutions. Thus a solution preserving $31$ supersymmetries, a BPS
preonic solution, is not allowed.

This completes the proof of the absence of BPS preonic,
$\nu=31/32$ supersymmetric bosonic solutions in type II
supergravities {\it i.e.}, in the {\it classical} approximation to
the type II string theories.

\section{The case of D=11 supergravity}

It is known that the $D=10$ type IIA supergravity can be obtained
by dimensional reduction from $D=11$ supergravity {\it i.e.}, its
solutions can be identified with solutions of $D=11$ supergravity
that are independent of one of the coordinates.
 In particular, the type IIA dilatino
$\check{\chi}^{\check{\alpha}}$, Eq. (\ref{IIADIL}), originates
from the 11-th component $\check{\psi}_{\#}^{\check{\alpha}}$ of
the $D=11$ gravitino
$\check{\psi}_{\check{\mu}}^{\check{\alpha}}=(\check{\psi}_{{\mu}}^{\check{\alpha}},
\check{\psi}_{\#}^{\check{\alpha}})$; schematically,
\begin{eqnarray}\label{chi=psi11}
 \check{\chi}^{\check{\alpha}}=\check{\psi}_{\#}^{\check{\alpha}}\; . \qquad
\end{eqnarray}
The type IIA supersymmetry transformations can also be obtained
from those of $D=11$ by dimensional reduction . This implies, in
particular, that the IIA $M$-matrix (\ref{M=IIA}) comes from the
eleventh component of the $D=11$ generalized connection;
schematically,
\begin{eqnarray}
\label{M=t11}
M_{\check{\beta}}{}^{\check{\alpha}} = {(\omega+\check{t})}_{\#\;
\check{\beta}}{}^{\check{\alpha}} \; . \qquad
\end{eqnarray}

This observation provides a starting point to probe the existence
of BPS preonic solutions in $D=11$ supergravity or, more
precisely, among the purely bosonic solutions of the classical
$D=11$ supergravity \cite{CJS78}. It was shown in \cite{JG+SP02}
that the existence of $k$ Killing {\it spinors} ($k=31$ for
preonic solutions) implies the existence of $k(k+1)/2$ Killing
{\it vectors},
\begin{eqnarray}
\label{KillIJ}
K^{\check{a}}_{IJ}:= \check{\epsilon}^{\;\check{\alpha}}_I
\Gamma^{\check{a}}_ {\check{\alpha}\check{\beta}}
\check{\epsilon}^{\;\check{\beta}}_J \; ,  \qquad
\end{eqnarray}
such that both the metric and the field strength $F_4=dA_3$ of the
three-form  gauge field $A_3$ are invariant under `translations'
along the directions of $K^{\check a}_{IJ}$,

\begin{eqnarray}
\label{Kill2IJ} \delta_{K_{IJ}}g_{{\check\mu}{\check\nu}}=
2D_{(\check{\mu}} K_{\check{\nu})IJ} =0 \; , \qquad
\delta_{K_{IJ}} F_4:= {\cal L}_{K_{IJ}} F_4=0 \; .
\end{eqnarray}
This actually implies that any supersymmetric solution of $D=11$
can be considered (at least locally) as a solution of $D=10$ type
IIA supergravity lifted (`oxidized') to $D=11$. Thus, because of
the above negative result for the existence of preonic solutions
in type IIA supergravity, the only remaining possibility to have
BPS preonic solutions in the D=11 case requires that they result
from the `oxidization' of a less supersymmetric solution of the
$D=10$ type IIA supergravity.

If the lifting to $D=11$ has to produce more supersymmetries, we
need that one or more Killing spinors
$\epsilon^{\check{\alpha}}_{\tilde{I}}$ have non-vanishing
derivative in the direction of a Killing vector, schematically,
$\partial_{\#}\epsilon^{\check{\alpha}}_I \not=0$. In this way,
the set of $D=11$ Killing equations
 ${\cal D}_{\#}\epsilon^{\check{\alpha}}_I:=
D_{\#}\epsilon^{\check{\alpha}}_I -
\epsilon^{\check{\beta}}_I\check{t}_{\#\check{\beta}}{}^{\check{\alpha}}=0$
will no longer reduce (see Eq. (\ref{M=t11})) to the algebraic
equation (\ref{KEqDIL}). As a result, the arguments from the
discussion of the type IIA case would not apply in $D=11$ to
exclude the existence of a preonic solution.

A Killing spinor $\epsilon^{\check{\alpha}}_J$ can be
characterized \cite{JG+SP02} by means of three differential forms:
a Killing vector one-form $K_{1\; JJ}:=
e_{\check{a}}K^{\check{a}}_{JJ}$, a two-form $\Omega_{2\; JJ}$ and
a five-form $\Sigma_{5\; JJ}$. These forms are the diagonal
elements of the symmetric bilinear matrix forms with tensorial
components defined in Eq. (\ref{KillIJ}) and by
\begin{eqnarray}\label{2,5-IJ}
\Omega^{\check{a}_1\check{a}_2}_{IJ}:=
\check{\epsilon}^{\;\check{\alpha}}_I
\Gamma^{\check{a}_1\check{a}_2}_ {\check{\alpha}\check{\beta}}
\check{\epsilon}^{\;\check{\beta}}_J \; ,  \qquad
\Sigma^{\check{a}_1\ldots \check{a}_5}_{IJ}:=
\check{\epsilon}^{\;\check{\alpha}}_I \Gamma^{\check{a}_1\ldots
\check{a}_5}_ {\check{\alpha}\check{\beta}}
\check{\epsilon}^{\;\check{\beta}}_J \; .  \qquad
\end{eqnarray}

The independence of a Killing spinor on a coordinate $x^{\#}$ in
some direction would also imply the independence of its associated
Killing vector $K_{JJ}$ (Eq. (\ref{KillIJ})), of the two-form
$\Omega_{JJ}$ and of the five-form $\Sigma_{JJ}$ (Eq.
(\ref{Kill2IJ})) on that direction. As the direction $x^{\#}$
should be characterized by one of the Killing vectors, the result
of \cite{JG+SP02}, stating that  ${\cal L}_K\Omega_2=0$ and ${\cal
L}_K\Sigma_5=0$, implies the independence of the two- and the
five-form on $x^{\#}$. However, the Lie derivative of a Killing
vector with respect to another Killing vector, ${\cal L}_K
K^\prime_1$, may still be nonzero when there are two or more
Killing vectors. Thus, at present we cannot conclude that all
Killing spinors $\epsilon^{\check \alpha}_I$ are independent of
$x^\#$ so that, albeit rather exotic, the possibility of a $\nu=
31/32$ supersymmetric solution in $D=11$ supergravity remains
open.

\section{Could preonic BPS solutions still exist?}

The established absence of preonic solutions in type II
supergravities, {\it i.e.} for the {\it classical} approximations
to type II string theories, does not preclude the preonic
conjecture of \cite{BPS01}. At the time it was made, solutions
preserving more than $16$ out of the $32$ supersymmetries were not
known except for the fully supersymmetric ones (see
\cite{FiPa03}). It was already mentioned in \cite{BPS01} that a
kind of `BPS preon conspiracy' could produce that only composites
of some number of preons (but not the preons themselves) could be
found (`observed') as supergravity solutions.

On account of the fundamental role played by preons in the
classification of BPS states \cite{BPS01}, it is tempting to
speculate that the fact that type II supergravities do not have
preonic solutions rather points out at a need for their
modification. The most natural refinement to try is to take into
account stringy, $(\alpha^\prime)^3$-corrections to the
supergravity equations and to the supersymmetry transformation
rules of the supergravity fields. Preonic solutions in a `stringy
corrected' type IIA supergravity would be allowed if  the
corrections modified Eqs.(\ref{dP=allV}) by adding some terms
$\propto \sigma_{abcdf}$ and $\propto \tilde{\sigma}_{abcdf}$.
Schematically, the `required' modification would have to be of the
form
\begin{eqnarray}
\label{dP=allVq?} a{\lambda}^1_{{\alpha}} {\lambda}^1_{\beta} -
{1\over 2} \sigma^a_{{\alpha}\beta}D_a \Phi =0 \quad &\mapsto&
\qquad a{\lambda}^1_{{\alpha}} {\lambda}^1_{\beta} - {1\over 2}
\sigma^a_{{\alpha}\beta}D_a \Phi = Q^-_{abcde}
\sigma^{abcde}_{{\alpha}\beta} \; ,   \qquad \nonumber
\\ a{\lambda}^{2{\alpha}} {\lambda}^{2\beta} - {1\over 2}
\tilde{\sigma}^{a\, {\alpha}\beta}D_a \Phi = 0 \quad &\mapsto&
\qquad a{\lambda}^{2{\alpha}} {\lambda}^{2\beta} - {1\over 2}
\tilde{\sigma}^{a\, {\alpha}\beta}D_a \Phi = Q^+_{abcde}
\tilde{\sigma}^{abcde\;{\alpha}\beta} \; \qquad
\end{eqnarray}
for some $Q_{abcde}\propto (\alpha^\prime)^3$ (clearly, the
$\propto \sigma_a$ contribution could also be changed, but this is
not essential for the present schematic discussion). Such a
modification (\ref{dP=allVq?}) of Eq. (\ref{dP=allV}) might result
from the associated additions to the dilatino transformation rules
(\ref{susyDIL}) (of the type $\propto \tilde{\sigma}_{abcdf}$ plus
other terms not essential for our discussion). In terms of the $M$
matrix, this modification would imply
\begin{eqnarray} \label{M=IIA+Q} M\; \mapsto \; M + \left(
\matrix{ 0 & Q^-_{abcde} \sigma^{abcde}_{{\alpha}\beta} \cr
{Q}^+_{abcde} \tilde{\sigma}^{abcde\; {\alpha}\beta} & 0}\right)=
M + Q^{\pm}_{abcde} \Gamma^{abcde} {1\over 2}(1\pm\Gamma^{11}) \;
.
\end{eqnarray}

Direct calculations of the stringy corrections
\cite{alphaprime,PVW2000,GrossWitten86} to the supersymmetry
transformation rules have been hampered by the lack of a covariant
technique to calculate higher order loop amplitudes in superstring
theory\footnote{Such a technique has been recently proposed in the
framework of Berkovits's pure spinor approach \cite{NBloops} to
the covariant description of the quantum superstring.}.
Nevertheless, bosonic string calculations allowed to find stringy
corrections to the Einstein equation \cite{GrossWitten86}. The
influence of these corrections on the supersymmetric vacua and
their relevance for their supersymmetric properties \cite{CFPSS86}
was used to find corrections to the gravitino supersymmetry
transformation properties. As the discussions of the
$\alpha^\prime$ modifications have also been extended to the
eleven-dimensional theory \cite{PVW2000,M-thQC}\footnote{The
contributions to the generalized connection ({\it i.e.} to the
supersymmetry transformation rules for the gravitino) were
calculated for a particular background and, then, conjectured to
hold in general \cite{M-thQC} on grounds of their universal
form.}, one can obtain the `corrected' transformation rules for
the type IIA dilatino\footnote{For the heterotic string case, the
simplest possible corrections to the ($N=1$) dilatino $\chi$
transformation rules (see Eq. (23) in \cite{CFPSS86}) consist in
modifying (`renormalizing') the dilaton $\Phi$ appearing in the
standard supersymmetry transformations.} by dimensional reduction
from those of the $D=11$ gravitino and thus derive the expression
of the matrix $M$ in Eq.(\ref{M=IIA}) that incorporates the
`stringy corrected' counterpart of Eq. (\ref{M=t11}).

In this perspective it looks promising that the $D=11$ generalized
connection $\check{t}_{\tilde{\mu}} = (\check{t}_{{\mu}},
\check{t}_{10})$ ({\it cf.} (\ref{susyDIL}); here
$\tilde{\mu}=(\mu;\#)$ = $0,\ldots 9; 10$) considered in
\cite{M-thQC}, contains the terms
$\check{Q}^{{\tilde{\mu}}_1\ldots
\tilde{\mu}_6}\Gamma_{\tilde{\mu}_1\ldots \tilde{\mu}_6}$; their
dimensional reduction would produce, among others, the
contribution $\check{Q}^{\mu_1\ldots \mu_5\,
10}\Gamma_{{\mu}_1\ldots {\mu}_5} \Gamma^{11}$ which is of the
needed type, see Eq. (\ref{M=IIA+Q}) (the
$\Gamma^{10}\equiv\Gamma^{\#}$ in $D=11$ is the $\Gamma^{11}$ in
$D=10$ ).

To summarize, although it has been shown that a $\nu=31/32$
preonic solution is not allowed in the {\it classical} type II
supergravities (in \cite{GGPR06} for type IIB and here for type
IIA), a conclusive analysis with quantum stringy corrections,
providing a more precise description of string/M-theory, remains
to be done. If preons were found to exist when quantum corrections
are taken into account, it would be only natural on account of
their special role as the `quarks of M-theory' \cite{BPS01}
\footnote{Let us recall \cite{BPS01} that the potential relevance
of BPS preons derives from their quark-like role in the
classification of BPS states according to the number of preserved
supersymmetries: all BPS states of M-theory preserving $k$
supersymmetries can be considered as composites of $32-k$ BPS
preons (the statement is true for any $D$ with 32 replaced by the
corresponding spinor dimension).}. Preons would only be `seen' by
looking at the `quantum solutions' of string theory, an
approximation of which is provided by supergravity with stringy
corrections.

As far as the study of `classical' supergravity is concerned, the
natural next step is to clarify the level of the mentioned `preon
conspiracy' \cite{BPS01} in the classical $D=10$ supergravity {\it
i.e.}, whether it is possible to find two-preonic $\nu=30/32$
supersymmetric solutions, preserving all but two supersymmetries,
or whether the `counterpart' of the colourless quark states in the
case of preons should include no less than four preonic
constituents corresponding to the  highest non-fully
supersymmetric states up to now found, the $\nu=28/32$ states of
the IIB case \cite{BeRo03}.

As for $D=11$ supergravity, although we have not been able to
reach a definite conclusion on the existence of $\nu=31/32\,$
supersymmetric solutions, we have presented here their
characteristic properties: such a $D=11$ BPS preonic solution
should have Killing directions, both for the metric $g$ and the
gauge field strength $F_4$, such that at least one of its $31$
Killing spinors depends on the coordinates corresponding to these
directions.

Finally, we conclude by mentioning that all searches for preonic
solutions, including this one, have been concerned with purely
bosonic solutions, a restriction that does not follow from
\cite{BPS01}.

\bigskip
\bigskip

{\bf Acknowledgments}. The authors wish to thank J. Gauntlett and
D. Sorokin for useful discussions. This work has been partially
supported by research grants from the Ministerio de Educaci\'on y
Ciencia (FIS2005-02761) and EU FEDER funds, the Generalitat
Valenciana (ACOMP06/187, GV05/102), the Ukrainian State Fund for
Fundamental Research (N383), the INTAS (2005/2006-7928) and by the
EU `Forces Universe' network (MRTN-CT-2004-005104).

\end{document}